\documentclass[aps,pra,twocolumn,showpacs,floatfix,epsfig,epsf]{revtex4}
\usepackage{amsmath}
\usepackage{amssymb}
\usepackage{graphicx}
\usepackage{subfigure}
\usepackage{natbib}
\usepackage{color}
\usepackage{epstopdf}
\usepackage{keyval}
\usepackage{trig}
\usepackage{revsymb}
\usepackage{hyperref}
\begin{document}

\title{Quantum processing by adiabatic transfer through a manifold of dark states}

\author{Santosh Kumar}
\author{Deepak Kumar}
\email{dk0700@mail.jnu.ac.in}
\affiliation{School of Physical Sciences, Jawaharlal Nehru University, New Delhi 110067, India}
\date{31 January 2012}

\begin{abstract}
We consider a network whose nodes are electromagnetic cavities, each coupled to a single three-level atom. The nodes are connected by optical fibers. Each atom is addressed by a control laser, which along with the cavity field drives atomic transitions. The network can be in the form of chain or two and three dimensional arrays of $N$-cavities connected by $N_B$ fibers. Following the work on two-cavity system by Pellizzari, we find that under certain conditions, the system possesses two kinds of dark states. The first kind are $N$ states corresponding to atomic excitations at each node and these are our logical states for quantum processing. The second kind are $N_B$ degenerate dark states on pairs of sites connected by a fibre. By manipulating intensities and phases of control lasers on the cavities, one can pass adiabatically among these dark states due to their degeneracy. This network operates as a $N$-level quantum system in which one can generate computationally useful states by protocols of external controls. We obtain numerical results for small chains and lattices to demonstrate some quantum operations like the transport of states across the array, generation of W-states and Fourier-like states. We also discuss effects of dissipation and limitations of the model.

\end{abstract}


\maketitle

A scalable quantum network is one of the most desired goals for quantum information processing \cite{NielsonChuang,BouwmeesterBook,Kimble08,DiVincenzo}. A quantum network consists of nodes which are connected by communication channels. The nodes contain qubits and allow for storage and local processing of quantum information. The communication channels are used to transfer quantum states as well as to create quantum entanglement among nodes. In the past two decades several schemes have been proposed to realize quantum networks, which exploit a variety of physical principles viz optical processes for polarization qubits, ion arrays in laser traps, electromagnetic cavities with trapped atoms, Josephson junction arrays, spin and charge qubits in quantum dots and schemes using NMR \cite{Zoller05, Qist,Blatt08}.%, Scala 
The merit of a scheme is guided by several criteria, chief among them being: (i) the ease of physical implementation (ii) robustness against noise and physical defects (iii) scalability to larger networks and (iv) simpler and flexible controls to guide quantum processes for desired ends.

Here we consider a network whose nodes are high-finesse electromagnetic cavities, each coupled to a single atom which has three suitable levels for quantum processing. The cavities are connected by optical fibers. Each atom can be addressed by a control laser. Networks based on cavity QED have been considered by a number of workers in the past, as this system exploits the long term memory and ease of processing of atomic qubits along with efficient communication with photons \cite{Cirac97,Pellizzari97,vanEnk99,Browne03,Clark03,Duan04,Agarwal04,Serafini06,Chen07,Song07,Lu08,Zhou09,DuanMonroe10,Zhong11}. Experimentally also there has been a remarkable progress which makes these networks amenable to implementation in a variety of ways exploiting besides conventional, technologies like optical microcavities, nano-fibre cavities \cite{Vahala03,Spillane05,Hunger10,Hakuta11}. Our proposal is derived from a scheme proposed by Pellizzari \cite{Pellizzari97} for a system of two cavities coupled by a fibre. This scheme achieves quantum state transfer between two three-level atoms by manipulating intensities of the two control lasers. It uses adiabatic passage through a dark state discussed below. %Duan03, Bose99,

We show that similar dark states also exist in a network of $N$ cavities arranged in a lattice with nearest neighbor cavities connected by $N_B$ fibers. Our considerations apply to many kinds of lattice in one to three dimensions. The full quantum mechanics of the network is very complex, but due to conservation of an excitation number it can be divided into sectors. All our operations and considerations are restricted to a low excitation sector. In this subspace, one finds that there are two kinds of dark states. The first kind are $N$ states that occur on each node involving only the atomic state. We shall use these as the computational states. The other kind are $N_B$ degenerate dark states in which states of atoms in the neighboring cavities and the photonic state in the connecting fibre are involved. By manipulating the intensities and phases of control lasers, one can pass adiabatically through these dark states and create all kinds of superpositions among the first kind of $N$ states. We show that this allows us to execute several computational tasks, though in a way different from the standard application of one-qubit and two-qubit gates.

In Fig. 1, we set up the notation by showing two nodes coupled by a fibre. The atomic levels are denoted by $\vert a_{0} \rangle$, $\vert a_{1} \rangle$ and $\vert b \rangle$. One cavity mode is coupled to the transition $\vert a_{1} \rangle \leftrightarrow \vert b \rangle$ and has a frequency $\omega_c$ . The external laser treated classically has a frequency $\omega_{l}$, and couples to the transition $\vert a_{0} \rangle \leftrightarrow \vert b \rangle$ with a Rabi frequency $\Omega$ which is a complex quantity. We first consider the ideal situation in which all dissipative effects including the spontaneous decay of the atomic levels are ignored. The dissipation will be considered later. The  Hamiltonian of a single atom-cavity system is $H=H_{0}+H_{I}$, with
\begin{eqnarray}
H_0 &=& \hbar \omega_c \left( C^{\dagger} C + \frac{1}{2} \right)  + \displaystyle\sum_{\alpha= a_{0}, a_{1},b} \hbar \omega_{\alpha} \vert \alpha \rangle \langle \alpha \vert, \\
H_{I} &=& \hbar g \left[ \vert a_{1} \rangle \langle b \vert C^{\dagger} + C \vert b \rangle \langle a_{1} \vert \right] - \hbar \left[ \Omega e^{-i\omega_{L} t} \vert b \rangle \langle a_{0} \vert \right. \nonumber \\&&  \left. + \Omega^{*}  e^{i \omega_{L} t} \vert a_{0} \rangle \langle b \vert \right],
\end{eqnarray}
where $C^{\dagger}$ is the creation operator for the cavity mode. We take the laser to be detuned from the atomic transition $\vert\Delta \vert$ $\gg \Omega (t), g$ with $\Delta = \omega_{l}-\omega_b +\omega_{a_0}$. The detuning satisfies the Raman condition $ (\omega_{c}-\omega_{l}) = (\omega_{a_{0}}-\omega_{a_{1}})$.

\begin{figure}[htbp]
\begin{center}
 \scalebox{0.4}{\includegraphics{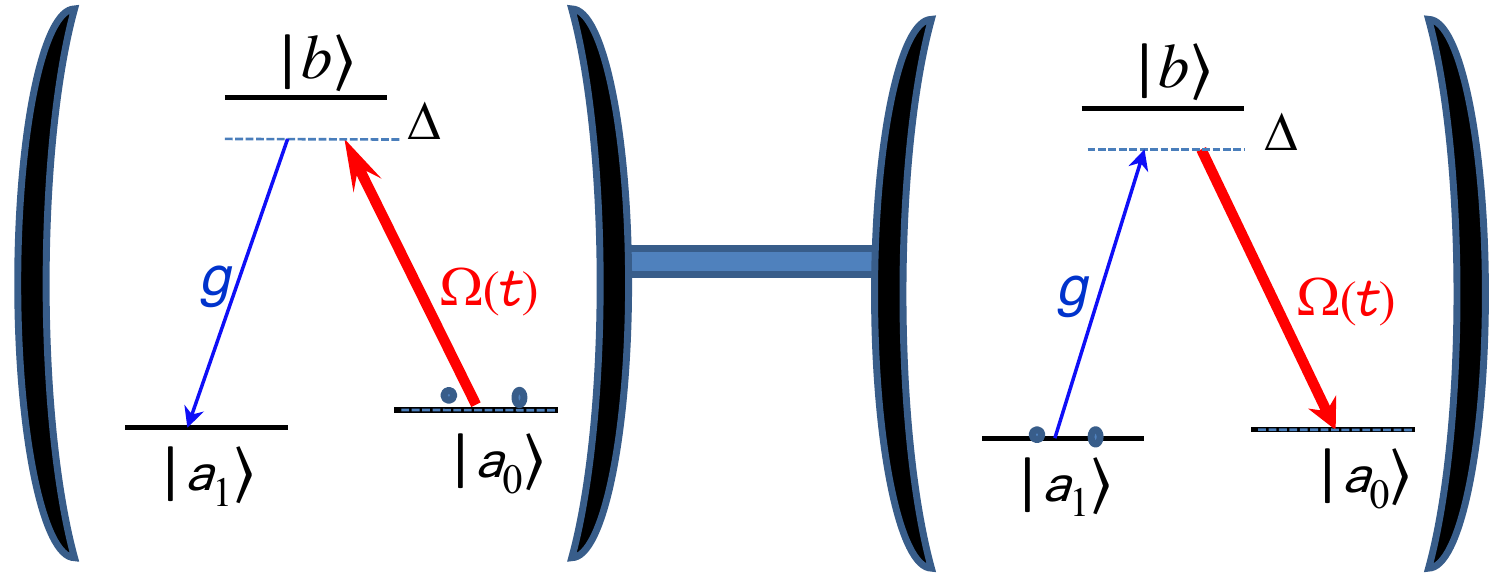}}
\end{center}
\caption{(color online) Two atom-cavity system connected by a fibre}
\end{figure}

Under these conditions of detuning, the residence time in level $b$ is small and one can adiabatically eliminate the excited state \cite{Pellizzari97}. This gives a two-level effective Hamiltonian for the system $H_{eff}$, given by
\begin{eqnarray}
 H_{eff} &=& \frac{\vert \Omega \vert^{2}}{\Delta} \vert a_{0} \rangle \langle a_{0} \vert + \frac{\vert g \vert^{2}}{\Delta}  C^{\dagger}C \vert a_{1} \rangle \langle a_{1} \vert  \nonumber \\
&+&\frac{g }{\Delta} \left[\Omega C \vert a_{0} \rangle \langle a_{1} \vert + \Omega^* C^{\dagger} \vert a_{1} \rangle \langle a_{0} \vert \right].\nonumber 
 \end{eqnarray}
The first two terms in $H_{eff}$ are diagonal and give Stark shifts of the two levels due the cavity and laser fields. For this scheme to work it is necessary to compensate the Stark shifts of the two levels by using other lasers which couple $a_{0}$ and $a_{1}$ non-resonantly to levels farther up \cite{Pellizzari97,Li10}. This leads to the following Hamiltonian for two fibre-coupled cavities
\begin{eqnarray}
%\label{Htwo}
 H = \sum^{2}_{j=1} \left[s_j C^{\dagger}_{j} \vert a_{1},1 \rangle \langle a_{0},0 \vert \right]
 + w (C_{1} + C_{2})X^{\dagger}_{12} +H.c.,
  \end{eqnarray}
where $s_j= \frac{g_{j} \Omega_{j}} {\Delta_{j}}$, $w$ is the cavity-fibre coupling strength and  $X^{\dagger}_{12}$ is the photon creation operator in the linking fibre. We include only one mode of the fibre which is resonant to the cavity mode. As shown by Pellizzari \cite{Pellizzari97}, this system has the following `Dark States' which have zero energy by choice, $\omega_{a_0}=\omega_{a_1}=0$.
\begin{eqnarray}
\vert p_1\rangle &=& \vert a_{0},0\rangle_{1} \vert a_{1},0\rangle_{2} {\vert 0\rangle }_f;\; \vert p_2\rangle = \vert a_{1},0\rangle_{1} \vert a_{0},0\rangle_{2} {\vert 0\rangle }_f\nonumber \\
\vert D_{1,2}\rangle &=& \big[\frac{1}{s_{1}} \vert a_{0},0\rangle_{1} \vert a_{1},0\rangle_{2} {\vert 0\rangle }_f
- \frac{1}{w} \vert a_{1},0\rangle_{1} \vert a_{1},0\rangle_{2} {\vert 1\rangle }_f \nonumber \\ &&
+ \frac{1}{s_{2}} \vert a_{1},0\rangle_{1} \vert a_{0},0\rangle_{2} {\vert 0\rangle }_f \big] 
  \end{eqnarray}
The first two, to be termed as P-states, are atomic states decoupled by the laser interaction in a trivial way. The second one is regarded as a dark state as it involve no states with cavity photons. This state is used in adiabatic transfer of atomic states \cite{Pellizzari97} from cavity 1 to cavity 2 by varying laser intensities through parameters $s_1$ and $s_2$.

Now we go on to consider lattices of coupled cavities as described above. Though these considerations apply to chains and other lattices in two and three dimensions, here we shall present explicit results for linear chain and square lattice. So we write below the effective Hamiltonian generalized from the two-cavity case, for a square lattice.
\begin{eqnarray}
 H &=& \sum^{N}_{i=1} \left[s_{i}(t) \vert a_{1},1 \rangle_{ii} \langle a_{0},0 \vert  C^{\dagger}_{i}
  + w (C_{i} + C_{i+\delta_{x}})X^{\dagger}_{i}  \right. \nonumber \\&&  \left.
                + w (C_{i} + C_{i+\delta_{y}})Y^{\dagger}_{i} \right]+H.c.,
\end{eqnarray}
 where $i+\delta_{x}$ and $i+\delta_{y}$ denote the right and upper neighbours of the site $i$ respectively. $X^{\dagger}_{i} (Y^{\dagger}_{i})$ create photons in fibers connecting sites $i$ and $i+\delta_{x} (i+\delta_{y})$.

The basis set for the wave function is written in the following notation,
$ \prod^{N}_{i=1} \vert a_{\mu_{i}},n_{i} \rangle_{i} \prod^{B}_{b=1}\vert m_{b} \rangle$,
where $\mu_{i}$ takes values 0 and 1, $n_{i}$ denotes the photon number in $i^{th}$ cavity and $m_{b}$ denotes the photon number in the $b^{th}$ fibre. The quantity $M = \sum_i\left[(1-\mu_i) + n_i \right]+ \sum_b m_{b}$ is conserved by the Hamiltonian. We shall work in the M = 1 sector, which has either zero photons, one atom in  $a_0$ state and the rest in $a_1$ state or with one photon and all atoms in $a_1$ state.  Thus only the following basis functions occur, which we denote by a condensed notation. $\vert p_i \rangle = \vert a_{1},0 \rangle_{1}\vert a_{1},0 \rangle_{2}...\vert a_{0},0 \rangle_{i}...\vert a_{1},0 \rangle_{N} \vert 0 \rangle_{1}...\vert 0 \rangle_{b}...\vert 0 \rangle_{B}$; $\vert q_i \rangle = \vert a_{1},0 \rangle_{1}\vert a_{1},0 \rangle_{2}...\vert a_{1},1 \rangle_{i}...\vert a_{1},0 \rangle_{N} \vert 0 \rangle_{1}...\vert 0 \rangle_{b}...\vert 0 \rangle_{B}$;
$\vert f_{i+\delta_{\alpha}} \rangle = \vert a_{1},0 \rangle_{1}\vert a_{1},0 \rangle_{2}...\vert a_{1},0 \rangle_{N} \vert 0 \rangle_{1}...\vert 1 \rangle_{i+\delta_{\alpha}}...\vert 0 \rangle_{B}$, with $\alpha=x,y$. The general wave function in this sector can be written as
\begin{eqnarray}
\vert \Psi (t)\rangle &=& \sum_{i} [\left(A_{i} (t) \vert p_i \rangle + B_{i} (t) \vert q_i \rangle + F_{i+\delta_x} (t) \vert f_{i+\delta_{x}} \rangle  \right. \nonumber \\&&  \left. +  F_{i+\delta_y} (t) \vert f_{i+\delta_{y}} \rangle \right] 
\end{eqnarray}

The equation of motion are
\begin{eqnarray} 
i\frac{\partial}{\partial t} A_{i}(t) &=& s^{*}_{i}(t) B_{i}(t)  \nonumber\\
i \frac{\partial}{\partial t} B_{i}(t) &=& s_{i}(t) A_{i}(t) + w \left[ F_{i+\delta_x}  \right. \nonumber \\&&  \left. + F_{i-\delta_x} + F_{i+\delta_y} + F_{i-\delta_y} \right]  \nonumber\\
i\frac{\partial}{\partial t} F_{i+\delta_x}(t) &=&  w \left[ B_{i}(t) +  B_{i+\delta_x}(t) \right]  \nonumber\\
i \frac{\partial}{\partial t} F_{i+\delta_y}(t) &=&  w \left[ B_{i}(t) +  B_{i+\delta_y}(t) \right].  
\end{eqnarray}
We again look for `Dark states' corresponding to zero energy. It is easy to check that the state,
\begin{eqnarray} 
D_{i,i+\delta_{\alpha}}&=& \left[\frac{1}{s_{i}} \vert p_{i}\rangle - \frac{1}{w} \vert f_{i+\delta_{\alpha}}\rangle + \frac{1}{s_{i+\delta_{\alpha}}} \vert p_{i+\delta_{\alpha}}\rangle \right],
\end{eqnarray}
is an eigenstate of zero energy. Number of these states is equal to B.
We have here a manifold of degenerate `Dark states' termed as D-states. When the cavity-fibre coupling is large, $w >> \{s_i\}$, the dark state is largely a superposition of two atomic excitations. One would like to operate the network in this regime.

The key point of this paper is that by adiabatic passage through the D-states one can generate any linear combination of P-states and pass from one combination to another. The D-states are not orthogonal, so it is difficult to treat the dynamics analytically for large $N$. However their nature helps us devise protocols for control lasers which take us from an initial state to a target state. Here all operations are controlled externally, unlike most other schemes where a particular unitary transformation depends on fixing a precise time for the Hamiltonian evolution.

\begin{figure}[htbp]
\begin{center}
 \scalebox{0.34}{\includegraphics{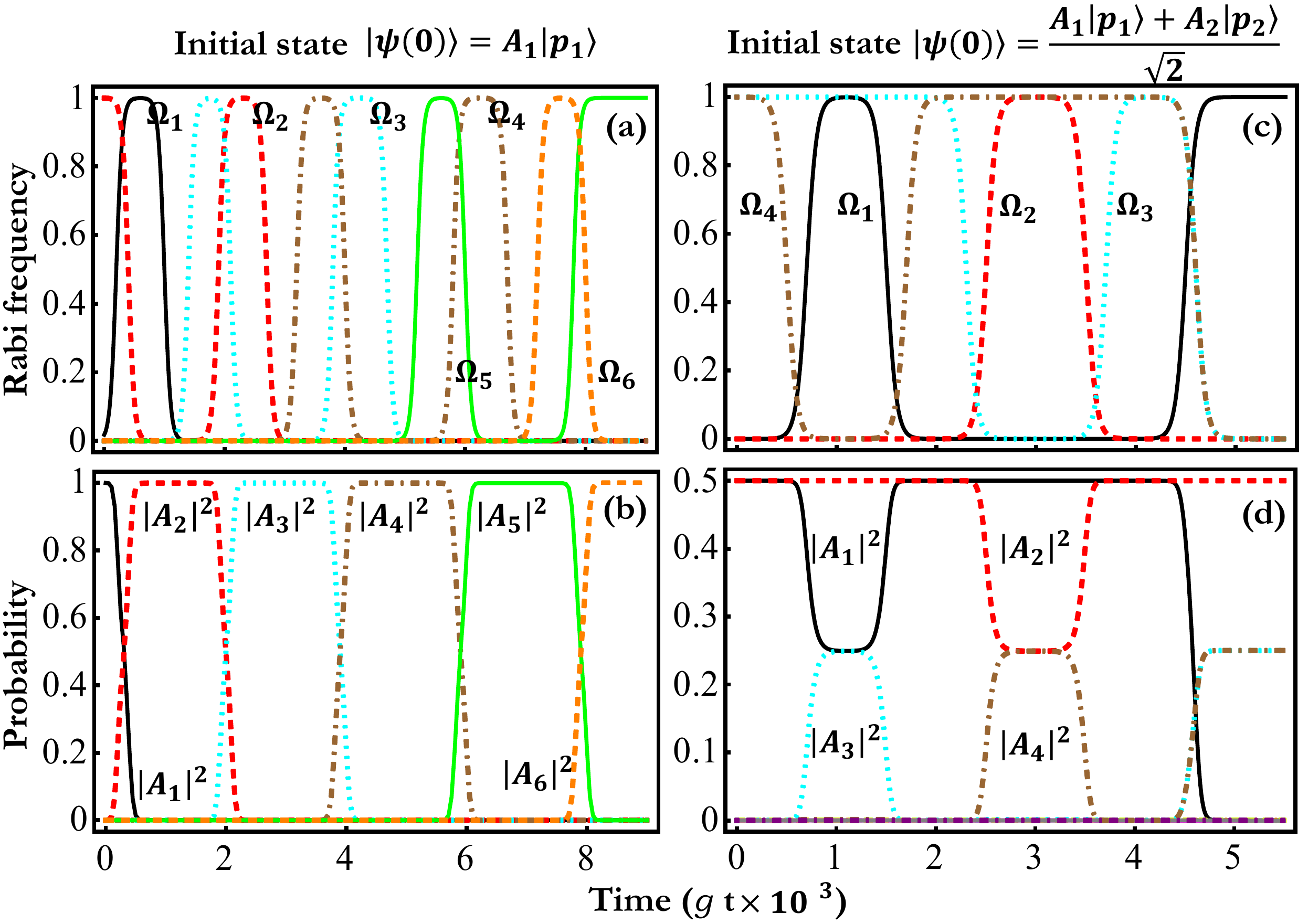}}
\end{center}
\caption{(color online) Successive transfer of the quantum state $a_0$ on nodes of a chain of six sites. Fig. (a) shows time variations of Rabi frequencies of control lasers and (b) the probabilities $\vert A_i\vert^2$ for states $\vert p_i \rangle$ exhibiting the transfer. Figs. (c) shows the Rabi-frequency protocol for generation of  superposed states on different sets of three nodes in a ring of four sites and (d) the corresponding probabilities. Parameters for the calculation are $\Delta = 3 g$, $w = 10 g$.}
\end{figure}
 
\begin{figure}[htbp]
\begin{center}
 \scalebox{0.34}{\includegraphics{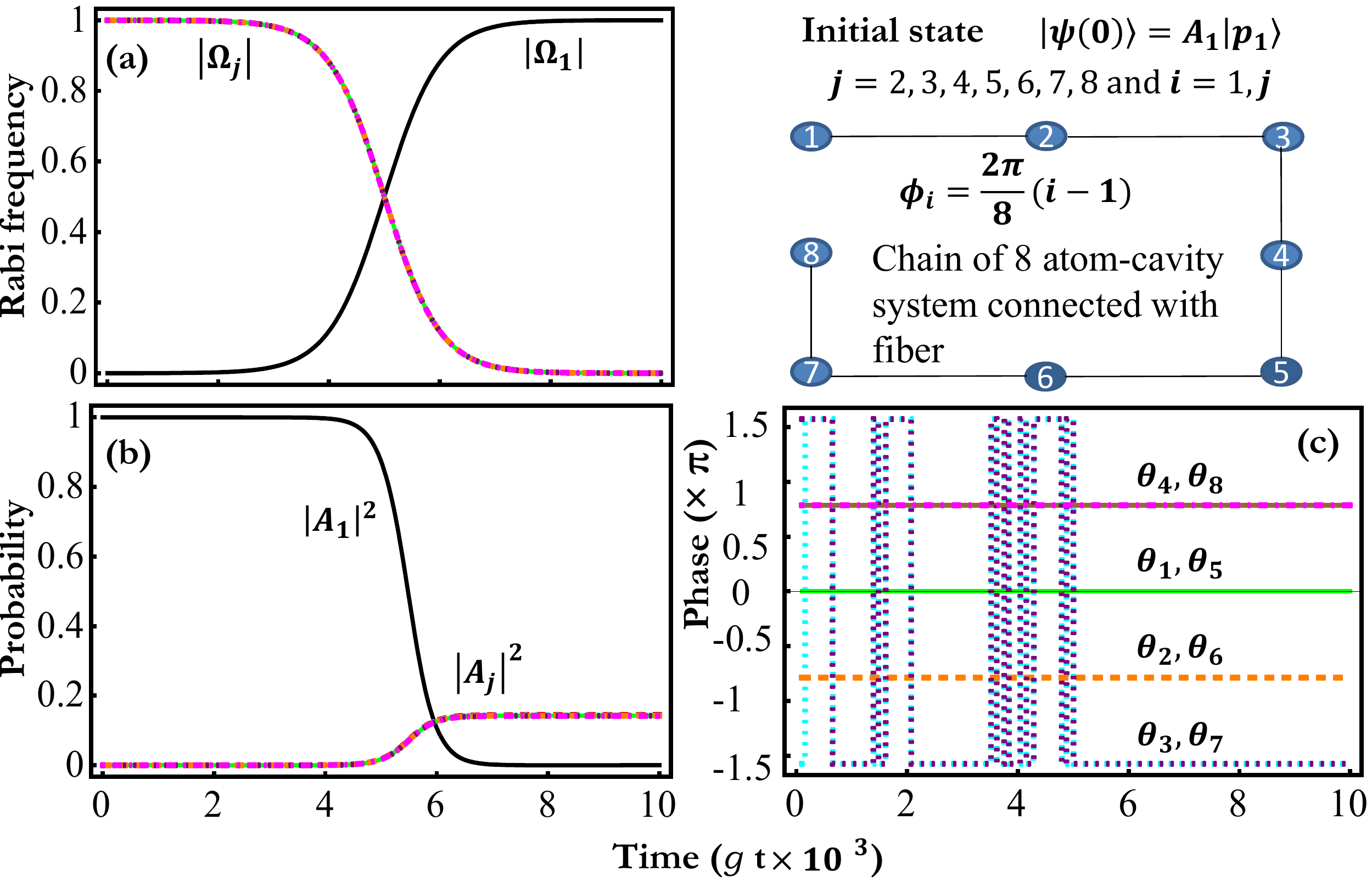}}
\end{center}
\caption{(color online) Generation of a Fourier-like state on a chain of 8 sites. (a) Time variations of the absolute Rabi frequency of control lasers. (b) Probabilities $\vert A_i\vert^2$ of atomic excitation $a_0$ on sites. They are eventually equal for sites 2 to 8. (c) Variation of phases with time at 8 sites. Note that eventual phases are equispaced. Calculation parameters are same as in Fig. 2. }
\end{figure}
\begin{figure}[htbp]
\begin{center}
 \scalebox{0.34}{\includegraphics{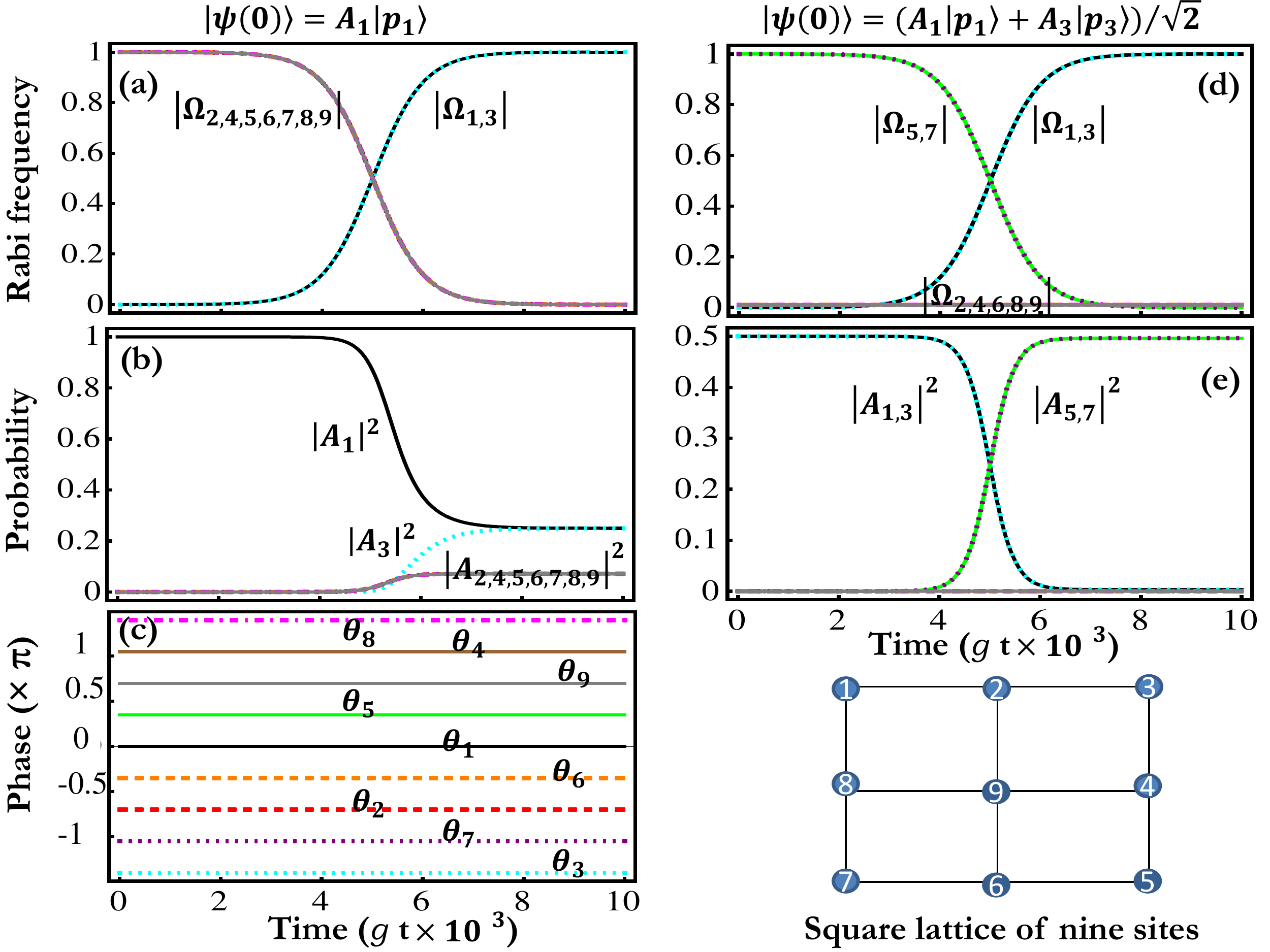}}
\end{center}
\caption{(color online) Generation of superposed states on a square lattice of nine sites.  Time variations of absolute Rabi frequencies of control lasers shown in (a) generate the state shown in (b) in which magnitudes of amplitudes $A_1$ and $A_3$ have one value and magnitudes of the other seven amplitudes have another value. (c) The phases of these amplitudes are equispaced. The Rabi frequency protocol shown in (d) transfers a two-node superposed state initially on nodes 1 and 3 to nodes 5 and 7 as shown by magnitudes of amplitudes in (e). Other parameters are same as in Fig. 2. }
\end{figure}

\begin{figure}[htbp]
\begin{center}
 \scalebox{0.34}{\includegraphics{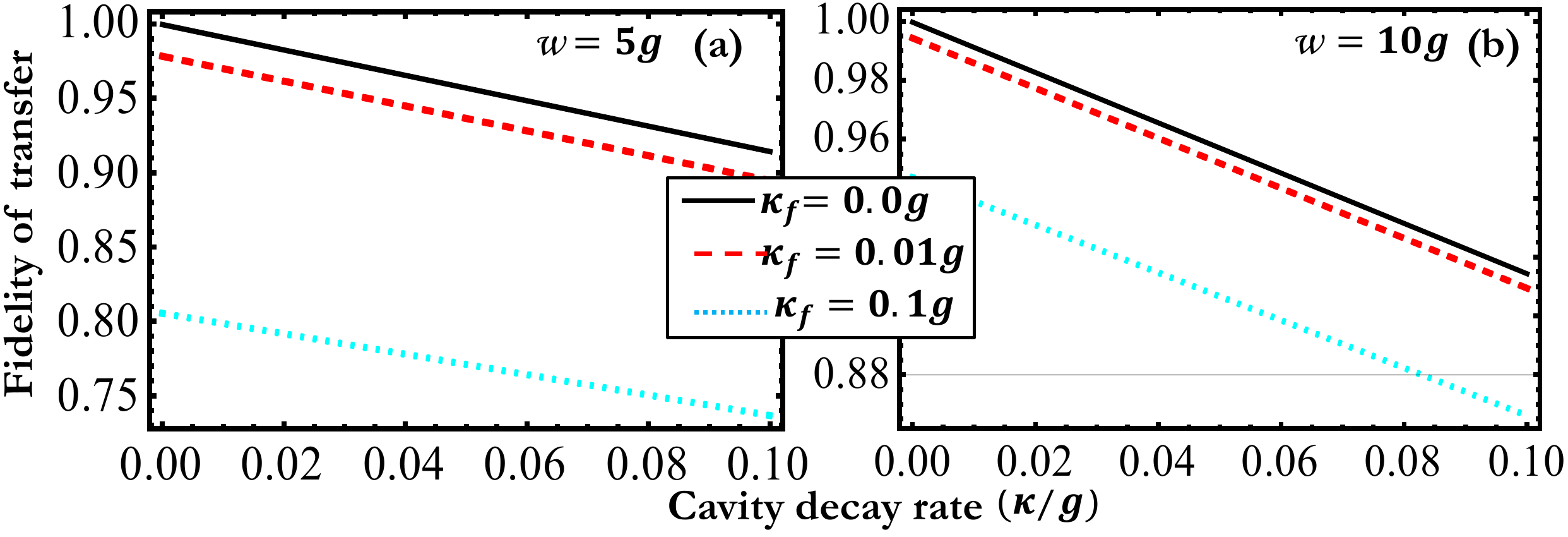}}
\end{center}
\caption{(color online) Effects of dissipation are shown through fidelity of transfer across a chain of four sites for two cavity-fibre couplings (a) $w$ = 5 $g$ and (b) $w$ = 10 $g$ and a range of cavity decay rates $\kappa$. Black solid line corresponds to $\kappa_{f}$ = 0, red dashed line is represents to $\kappa_{f}$ = 0.01 $g$ and  cyan dotted line is represents to $\kappa_{f}$ = 0.1 $g$ and $\gamma$ = 0.1 $g$.}
\end{figure}

Since this sector has a single atomic excitation ($a_1$ to $a_0$), it does not provide us with single node or two-node qubits as in conventional networks \cite{NielsonChuang}. On the other hand it offers us a highly manipulable $N$-state quantum system, in which several computationally useful states can be generated by simple interpretation. We can associate sites of the lattice with numbers $x$ from 1 to $N$. For parallel processing, the state used in algorithms like those of Shor \cite{Shor} or Grover \cite{Grover}  is: $\vert \psi \rangle = \frac{1}{\sqrt N} \sum_{x=1}^N \vert x \rangle $. For this state and its Fourier transform $\vert \vec Q \rangle = \frac{1}{\sqrt N} \sum_{x=1}^N \vert x \rangle \exp (i\vec Q.\vec R(x))$, where $\vec R(x)$ denotes the site site assigned $x$, protocols can be devised which seem easy to implement, as shown below.

We now present some typical results of numerical simulations of Eqs.(7). We use parameters that are commonly achieved in recent experimental setups \cite{vanEnk99,Spillane05,Hunger10,Trupke05}. The time variation of Rabi frequencies is of the form $\tanh r(t+t_0)$. In Fig. 2 we show a successive transfer of atomic excitation $a_0$ on the nodes of a chain of six sites. The protocol for the variation of Rabi frequencies are shown in Fig. 2(a), while the probabilities $\vert A_i\vert^2$ for states $\vert p_i \rangle$ are shown in Fig. 2(b). Fig. 2(c) takes an initial state $\frac {\vert p_1 \rangle + \vert p_2 \rangle}{\sqrt 2}$ to states of the form $\frac{\vert p_2 \rangle}{\sqrt 2}+\frac {\vert p_i \rangle + \vert p_j \rangle}{2}$ as shown by magnitudes of amplitudes in Fig. 2(d). In Fig. 3, we show the protocol (Fig. 3(a)) and results for the generation of states $\vert \psi \rangle$ and $\vert Q \rangle$ for a chain of eight nodes. From the initial state $\vert p_1 \rangle$ our protocol generates a state of equal amplitude on sites 2 to 8 as seen in Fig. 3(b) and equispaced phases $\theta_i$ as seen in Fig. 3(c). This requires taking Rabi frequencies in the form $\vert\Omega_n \vert e^{in \phi_0}$. If we suppress the phases, one obtains the $\vert \psi \rangle$ state. In Fig. 4, results on a lattice of 9 sites are exhibited. The protocol shown in Fig. 4(a) takes an initial state $\vert p_1 \rangle$ to a superposed state in which the magnitudes of $A_1$ and $A_3$ have one value and the remaining amplitudes are equal in magnitude with another value as seen in Fig. 4(b). The phases of these amplitudes are equispaced as seen in Fig. 4(c). The protocol of Fig. 4(d) achieves a transfer of a superposed state $\frac {\vert p_1 \rangle + \vert p_3 \rangle}{\sqrt 2}$ to $\frac {\vert p_5 \rangle + \vert p_7 \rangle}{\sqrt 2}$ as seen by the magnitudes of the amplitudes in Fig. 4(e). This demonstrates the control and the flexibility of this scheme.

The effect of dissipation is a key feature for the viability of quantum processing. The main sources of dissipation are: (i) Spontaneous decay of the level $\vert b \rangle$. (ii) loss of  photons from cavities mode. (iii) Decay of the photons in fibers. (iv) Motion of atoms. The effects (i) and (ii) are minimized in this scheme as the dark states used involve only states with zero number of photons in the cavities and no population in upper levels $\vert b \rangle_i$. However, this is contingent on doing operations adiabatically and in manner that the system does not transfer out of $M=1$ subspace.
The decay of photons in fibers should be minimized by having high quality short fibers and by increasing the cavity-fibre coupling, as that decreases the weightage of states with photons in fibers.

 Though we believe the above factors related to dissipative and nonadiabatic effects need a more refined analysis, following Pellizzari \cite{Pellizzari97}, we have numerically examined the role of dissipation in the framework of continuous measurement theory \cite{ZollerGardiner}. Here one assumes that the lost photons are being continuously monitored. This leads to an effective non-Hermitian Hamiltonian in which the parameters $s_i$ are modified to $\frac{g \Omega_i}{\Delta + i\gamma}$, where $\gamma$ denotes the spontaneous decay rate of the level b. Further one allows for loss of photons from the cavities and fibers by adding terms $-i\kappa \sum_j^N C^{\dagger}_j C_j -i\kappa_f \sum_b^{N_B} X^{\dagger}_b X_b $. Results are shown in Fig. 5 for the fidelity of transfer of a state across a four-node chain. Effects of $\gamma$ and $\kappa$ are seen to be reasonably small, but the effect of $\kappa_f$, the rate of fibre decay is considerable.

In conclusion, we have presented a proposal of a scalable network made up from an array of cavities connected with optical fibers in any dimension. By only external controls one obtains quantum state transfer and several superpositions of atomic states at the nodes of the network. We show that these states are computationally useful and allow possibilities of new algorithms. A preliminary analysis of effects of dissipation shows encouraging prospects.

We have benefited from discussions with Rupa Ghosh. SK and DK are supported respectively by fellowships from the Council of Scientific and Industrial Research and the Department of Atomic Energy, India.

\end{document}